\documentclass[twocolumn]{aastex62}

\usepackage{multirow}
\usepackage{amsmath}
\usepackage{longtable}
\usepackage{color}

\usepackage[encapsulated]{CJK}
\usepackage{ucs}
\usepackage[utf8x]{inputenc}

\newcommand{\vlsr}     {V_\mathrm{lsr}}
\newcommand{\vsys}     {V_\mathrm{sys}}
\newcommand{\vout}     {V_\mathrm{out}}

\newcommand{\pa} {\mathrm{P.A.}}

\newcommand{\HII}     {\mbox{H\,{\sc ii}}}
\newcommand{\FWHM}     {\mathrm{FWHM}}

\newcommand{\Iobs} {I_\mathrm{obs}}
\newcommand{\Imod} {I_\mathrm{mod}}

\newcommand{\h} {^{\mathrm{h}}}
\newcommand{\m} {^{\mathrm{m}}}
\newcommand{\s} {^{\mathrm{s}}}
\newcommand{\Jybeam}  {\mbox{Jy}~\mbox{beam}^{-1}}
\newcommand{\mJybeam}  {\mbox{mJy}~\mbox{beam}^{-1}}

\newcommand{\kms}	{\mbox{km s}^{-1}}

\newcommand{\kpc}	{\mbox{kpc}}
\newcommand{\yr}	{{\rm yr}}
\newcommand{\K}	{{\rm K}}

\newcommand{\au} {\mbox{au}}
\newcommand{\GHz} {\mbox{GHz}}

\shorttitle{Ionized Bipolar Outflow from G45.47+0.05}
\shortauthors{Zhang et al.}

\begin{document}

\title{Discovery of a Photoionized Bipolar Outflow towards the Massive Protostar G45.47+0.05}

\author{Yichen Zhang}
\affiliation{Star and Planet Formation Laboratory, RIKEN Cluster for Pioneering Research, Wako, Saitama 351-0198, Japan; yczhang.astro@gmail.com}

\author{Kei E. I. Tanaka}
\affiliation{Department of Earth and Space Science, Osaka University, Toyonaka, Osaka 560-0043, Japan}
\affiliation{ALMA Project, National Astronomical Observatory of Japan, Mitaka, Tokyo 181-8588, Japan}

\author{Viviana Rosero}
\affiliation{National Radio Astronomy Observatory, 1003 Lopezville Rd., Socorro, NM 87801, USA}

\author{Jonathan C. Tan}
\affiliation{Department of Space, Earth \& Environment, Chalmers University of Technology, SE-412 96 Gothenburg, Sweden}
\affiliation{Department of Astronomy, University of Virginia, Charlottesville, VA 22904-4325, USA}

\author{Joshua Marvil}
\affiliation{National Radio Astronomy Observatory, 1003 Lopezville Rd., Socorro, NM 87801, USA}

\author{Yu Cheng}
\affiliation{Department of Astronomy, University of Virginia, Charlottesville, VA 22904-4325, USA}


\author{Mengyao Liu}
\affiliation{Department of Astronomy, University of Virginia, Charlottesville, VA 22904-4325, USA}

\author{Maria T. Beltr\'an} 
\affiliation{INAF -- Osservatorio Astrofisico di Arcetri, Largo E. Fermi 5, 50125 Firenze, Italy}



\author{Guido Garay}
\affiliation{Departamento de Astronom\'ia, Universidad de Chile, Casilla 36-D, Santiago, Chile}

\begin{abstract}
Massive protostars generate strong radiation feedback, which may help
set the mass they achieve by the end of the accretion
process. Studying such feedback is therefore crucial for understanding
the formation of massive stars. We report the discovery of a
photoionized bipolar outflow towards the massive protostar G45.47+0.05
using high-resolution observations at 1.3~mm with the Atacama
Large Millimeter/Submillimeter Array (ALMA) and at 7~mm with
the Karl G. Jansky Very Large Array (VLA). By modeling the free-free
continuum, the ionized outflow is found to be a photoevaporation flow
with an electron temperature of $10,000~\K$ and an electron number density of
$\sim1.5\times10^7~\mathrm{cm}^{-3}$ at the center,
launched from a disk of radius of $110~\au$. 
H30$\alpha$ hydrogen recombination
line emission shows strong maser amplification, with G45 being
one of very few sources to show such millimeter recombination line masers. The mass
of the driving source is estimated to be $30-50~M_\odot$ based on the
derived ionizing photon rate, or $30-40~M_\odot$ based on the
H30$\alpha$ kinematics.  The kinematics of the photoevaporated
material is dominated by rotation close to the disk plane, while
accelerated to outflowing motion above the disk plane.  The
mass loss rate of the photoevaporation outflow is estimated to be
$\sim(2-3.5)\times 10^{-5}~M_\odot~\yr^{-1}$.  We also found hints
of a possible jet embedded inside the wide-angle ionized outflow with
non-thermal emissions. The possible co-existence of a jet and a
massive photoevaporation outflow suggests that, in spite of the strong
photoionization feedback, accretion is still on-going.
\end{abstract}

\keywords{ISM: individual objects (G45.47+0.05), jets and outflows, HII regions --- stars: formation, massive}

\section{Introduction}
\label{sec:intro}

Massive stars dominate the radiative, mechanical and chemical feedback
to the interstellar medium, thus regulating the evolution of galaxies.
However, their formation process is not well understood.  One key
difference between high- and low-mass star formation is that massive
protostars become so luminous that they generate strong radiation
feedback, which potentially stops the accretion. For example,
radiation pressure was considered a potential barrier for massive star
formation \citep{wol87,yor02}. Photoionization is another important
feedback process, as massive protostars emit large amounts of Lyman
continuum photons that ionize the accretion flows to form
photoevaporative outflows driven by the thermal pressure of
$\sim10^4~\K$ ionized gas \citep{hol94}. However, theoretical
calculations and simulations have suggested that, in the core
accretion scenario these feedback processes are not strong enough to stop accretion
\citep{kru09,kui10,pet10,tan11}. \citet[]{Tanaka17} studied the
combined effects from various feedback processes in massive star
formation, including magnetohydrodynamical (MHD) disk winds, radiation
pressure, photoionization and stellar winds, finding that MHD winds
are dominant, while the others processes play relatively minor roles.
However, observational confirmation of such a theoretical scenario is still
difficult, due to the rarity and typically large distances of massive
protostars, especially the most massive type with strong radiation feedbacks.

The protostar G45.47+0.05 (hereafter G45; $d=8.4~\kpc$,
\citealt[]{Wu19}) has a luminosity of $\sim(2-5)\times10^5~L_\odot$
and a mass of $\sim20-50~M_\odot$ (\citealt[]{Debuizer17}), based on
infrared spectral energy distribution (SED) fitting 
(\citealt[]{ZT18}). G45 is associated
with an ultracompact (UC) $\HII$ region (\citealt[]{Wood89,Urquhart09,Rosero19}) and OH
and H$_2$O masers (\citealt[]{Forster89}).  Molecular outflows are
seen in HCO$^+$($1-0$), with blue-shifted emission to the north and
red-shifted emission to the south, 
consistent with the elongation of radio
continuum emission (\citealt[]{Wilner96}).
Infrared observations at $10-40~\mu$m show extended emission offset to the
north of the UC$\HII$ region (\citealt[]{Debuizer05,Debuizer17}),
which may come from the near-facing (blue-shifted) outflow cavity.
The offset between the infrared emission and the UC$\HII$ region may be due
to the high infrared extinction towards the protostar, suggesting dense
molecular gas surrounds the source.

\section{Observations}
\label{sec:obs}

ALMA 1.3 mm observations were performed on 2016 April 24,
2016 September 4 and 2017 November 2 with C36-3, C36-6 and C43-9
configurations (hereafter C3, C6 and C9), 
with on-source integration times of 3.6, 7.3 and 17.7~min, respectively
 (project IDs: 2015.1.01454.S; 2017.1.00181.S).
J1751+0939, J2025+3343 and J2000-1748 were used for bandpass
calibration; Titan, J2148+0657 and J2000-1748 were used for flux
calibration; and J1922+1530, J1914+1636 and J1922+1530 were used as
phase calibrators. 
The data were calibrated and imaged in CASA
(\citealt{McMullin07}).  Self-calibration
was performed for the three configuration data separately
using the continuum data (bandwidth of 2~GHz), and applied to the line
data.  Images are made with the CASA task tclean using robust
weighting (\citealt[]{Briggs95}) with the robust parameter of 0.5.  The synthesized beam sizes of
different configurations are listed in Figure
\ref{fig:contmap}.  Here we focus on continuum and H30$\alpha$ line
(231.90093~GHz) data, deferring analysis of molecular lines to a
future paper.

The VLA 7~mm observation was performed in 2014 with A
configuration and two 4-GHz basebands centered at 41.9 and 45.9~GHz
(project ID: 14A-113).  Details of the VLA data is summarized by
\citet[]{Rosero19}.

\section{Results}
\label{sec:results}

\subsection{1.3 and 7~mm continuum}
\label{sec:continuum}

\begin{figure*}
\begin{center}
\includegraphics[width=0.95\textwidth]{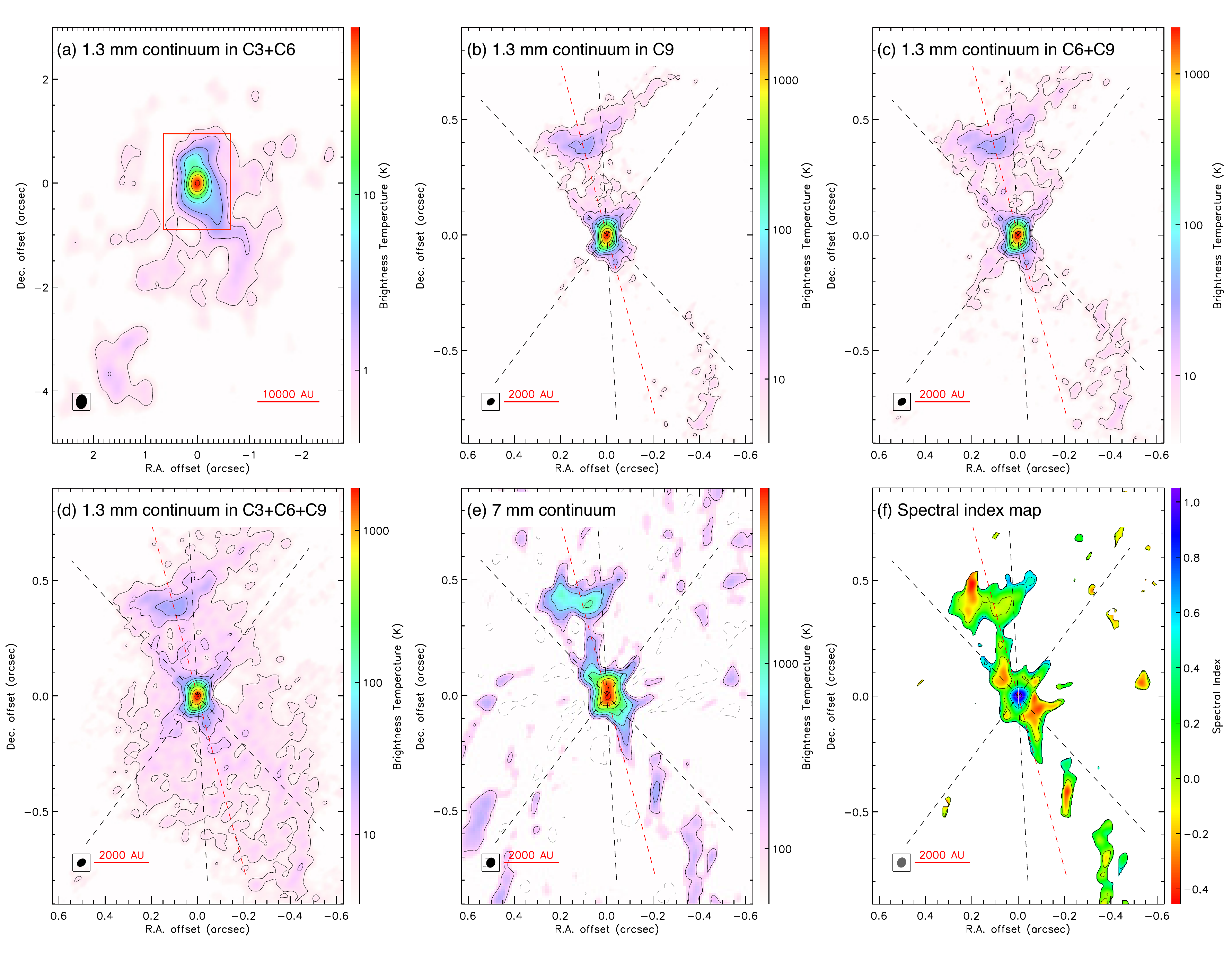}\\
\caption{
{\bf (a)$-$(d):} ALMA 1.3~mm continuum map of G45 observed in the
C3+C6 configuration (panel a), C9 configuration (panel b), C6+C9
configuration (panel c), and C3+C6+C9 configuration (panel d).  The
synthesized beams in these images are $0.28''\times0.21''$
($\pa=-6.6^\circ$, panel a), $0.038''\times0.026''$
($\pa=-53.9^\circ$, panel b), $0.042''\times0.028''$
($\pa=-53.6^\circ$, panel c), and $0.043''\times0.029''$
($\pa=-53.4^\circ$, panel d).  The contour levels are
$5\sigma\times2^n$ ($n=0,1,...$), with $1\sigma=0.13~\K$
($0.33~\mJybeam$, panel a), $1\sigma=1.1~\K$ ($0.049~\mJybeam$, panel
b), $1\sigma=0.91~\K$ ($0.047~\mJybeam$, panel c), and
$1\sigma=0.94~\K$ ($0.052~\mJybeam$, panel d).  {\bf (e):} VLA 7~mm
continuum map with solid contours at $5\sigma\times2^n$ ($n=0,1,...,6$,
$1\sigma=25~\K$ or $0.067~\mJybeam$),
and dashed contours at $-5\sigma$ and $-10\sigma$. 
The synthesized beam is
$0.045''\times0.038''$ with $\pa=-24.2^\circ$.  {\bf (f):} Map of
continuum spectral index between 1.3~mm (C3+C6+C9) and 7~mm,
$\alpha_\nu=\log(I_{\nu_1}/I_{\nu_2})/\log(\nu_1/\nu_2)$, where
$\nu_1=44~\GHz$ and $\nu_2=234~\GHz$.  Only regions where both 1.3 and
7~mm emissions are $>5\sigma$ are included.  Here the 1.3~mm image is
restored with the same beam size as the 7~mm image.  The red rectangle in
panel a marks the region shown in panels b$-$f.  In panels b$-$f, the
black dashed lines indicate the outflow axis ($\pa=3^\circ$) and
projected half-opening angle ($40^\circ$), and the red dashed line
indicates a possible jet ($\pa=15^\circ$).  The R.A. and
Dec. offsets in all the panels are relative to the continuum peak
position from the ALMA C9 configuration data,
$(\alpha_{2000},\delta_{2000})=(19\h14\m25\s.678$,
$+11^\circ09\arcmin25\arcsec.567)$.}
\label{fig:contmap}
\end{center}
\end{figure*}

The 1.3~mm continuum emission appears concentrated within $5\arcsec$
($4.2\times10^4\:\au$) from the central source (Figure
\ref{fig:contmap}a). At low resolution (C3+C6 configuration), the
continuum emission is elongated in the north-south direction, with
extended structures towards the south. At high resolution (C9
configuration), the continuum shows an hourglass shape aligned in the
north-south direction (Figure \ref{fig:contmap}b), with a morphology
highly symmetric with respect to an axis with $\pa\approx 3^\circ$,
which is consistent with the direction of the HCO$^+$ outflow
(\S\ref{sec:intro}).  The apparent half-opening angle of this
hourglass is $\sim40^\circ$ (dashed lines in Figure
\ref{fig:contmap}b).  More extended emissions are recovered by
combining with more compact configuration data (panels c and d). 
The bipolarity can be still clearly seen in the combined images, and most
of the fainter extended emissions are inside of the outflow cavity to
the north and south.
The 7~mm continuum (panel e) has very similar
morphology as the high-resolution 1.3~mm continuum.
At both 1.3 and 7~mm, most of the compact
continuum emissions are concentrated within
$0.2\arcsec$ ($1.7\times10^3~\au$) from the central source. 
Another emission peak is seen $\sim0.4\arcsec$ to the north, 
which was first identified by \citet[]{Rosero19}.  Its emission appears to be more
extended and shows an arc-like shape facing away from the main source
in a direction consistent with the main outflow, suggesting that the
northern emission structure is part of the main outflow.

Figure \ref{fig:contmap}f shows the spectral index map derived using
the 1.3 and 7~mm continuum intensities,
$\alpha_\nu=\log(I_{\nu_1}/I_{\nu_2})/\log(\nu_1/\nu_2)$, where
$\nu_1=44$~GHz and $\nu_2=234$~GHz.  Note that the position of the 1.3~mm
emission peak is slightly offset from either the emission peak or
the center of symmetry of the 7~mm image (see Figure
\ref{fig:contmap}e).  Therefore, we manually shift the 7~mm image by 7
mas in the R.A. direction and 13 mas in the Dec. direction, so that
the 1.3~mm emission peak coincides with the 7~mm center of symmetry
in making the spectral index map (also see Figure \ref{fig:ffmodel}d).
The central region has spectral indices of $\alpha_\nu>0.5$,
indicating partially optically thick ionized gas ($\alpha_\nu=2$ for
completely optically thick free-free emission), while more
extended structures have spectral indices of $\alpha_\nu\approx0$,
consistent with optically thin free-free emission
(e.g., \citealt[]{Anglada18}).  Note that, for dust continuum emission,
the index is $\alpha_{\nu,\mathrm{dust}}=2$ in the optically thick
case and $3\lesssim\alpha_{\nu,\mathrm{dust}}\lesssim4$ in the
optically thin case (assuming a typical dust emissivity spectral index
of $1-2$). These suggest that in this source, even at 1.3~mm, 
the compact continuum emission may have a significant free-free contribution,
if not dominated by it. This is further supported by
the fact that the 1.3 and 7~mm continuum emissions coincide well with
the H30$\alpha$ hydrogen recombination line (HRL) emission (Figure
\ref{fig:H30alpha}a; see below).  Furthermore, the 1.3~mm peak
brightness temperature is $\sim2000~\K$, higher than that expected
from dust continuum (the average dust temperature within 1000~au from
a $\sim30\:M_\odot$ protostar is estimated to be several $\times
10^2\:\K$ from dust continuum radiative transfer (RT) simulations by
\citet[]{ZT18}) and the dust sublimation temperature
($\sim1600~\K$), but can be naturally explained by thermal emission
from ionized gas with a typical temperature of $10^4\:\K$ around
massive protostars.

However, the extended 1.3~mm continuum emission shown in the low-resolution
data should still be dominated by dust emission.
The total 1.3~mm continuum flux $>3\sigma$ within $5\arcsec$ from the
central source measured from the C3+C6 image is 0.77~Jy. The compact
structure within $0.9\arcsec$ has a total $>3\sigma$ flux of 0.34~Jy
measured from the C9 image, or 0.61~Jy from the C3+C6+C9 image, which
can be considered as an estimate (upper limit) of the free-free flux.
Their difference of 0.16 to 0.43 Jy can be used as an estimate (lower
limit) for the extended dust continuum flux of the source, which
corresponds to a gas mass of $(2.1-5.6)\times 10^2\:M_\odot$ assuming
a temperature of 30~K (typically assumed for molecular cores), or
$(0.95-2.6)\times 10^2~M_\odot$ assuming 60~K (suitable for within
$\sim 10,000~\au$ around a $20-50~M_\odot$ protostar; \citealt[]{ZT18}). 
Here we adopt a dust opacity of
$\kappa_\mathrm{1.3mm}=0.899~\mathrm{cm}^2~\mathrm{g}^{-1}$
(\citealt[]{Ossenkopf94}) and a gas-to-dust mass ratio of 141
(\citealt[]{Draine11}).  These mass estimates suggest that there is a
large mass reservoir for future growth of the massive protostar.

Inside the wide-angle hourglass outflow cavity, a narrower
jet-like feature is apparent in the 1.3~mm images
with $\pa\approx 15^\circ$ (red dashed
line in Figure \ref{fig:contmap}). Along this direction,
a structure separate from the outflow cavity walls extends
from the central source to the northern lobe. To the south, along the same direction, there is
an additional emission peak $\sim0.1\arcsec$ from the center, and
also a stream of extended emission. These are most clearly seen in
Figures \ref{fig:contmap}b and \ref{fig:contmap}c.
Along this direction, there are also some emission features
in the 7~mm image.  However, some of these may arise from strong
side-lobe patterns from incomplete uv sampling in the VLA observation.
As shown in Figure \ref{fig:contmap}f, some regions have negative
spectral indices of $-0.5<\alpha_\nu<-0.2$, inconsistent with pure
free-free emission or dust emission, indicating possible contributions
from nonthermal synchrotron emission, which is occasionally found
associated with massive protostellar jets (e.g.,
\citealt[]{Garay96,Carrasco10,Moscadelli13,Rosero16,Beltran16,Sanna19}).
Thus, detection of negative spectral indices is also consistent with a jet
being embedded inside the wide-angle ionized outflow.
Evidence for the coexistence of wide-angle outflows with 
collimated jets has been found in other massive young stellar sources, 
such as Cepheus A HW2, where water masers trace a wide outflow and free-free continuum
traces a collimated jet (\citealt[]{Torrelles11}).

\subsection{H30$\alpha$ line}
\label{sec:H30alpha}

\begin{figure*}
\begin{center}
\includegraphics[width=0.8\textwidth]{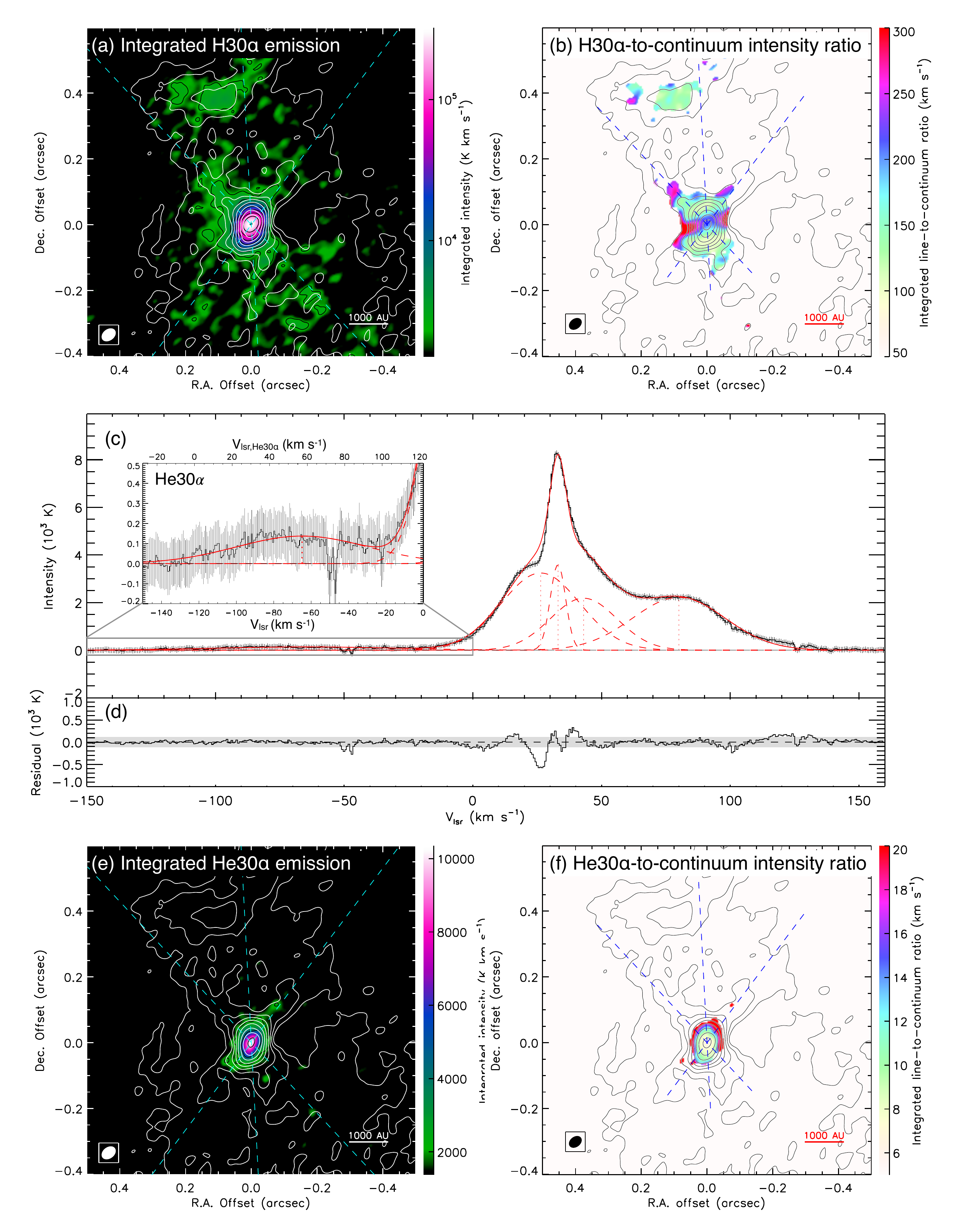}\\
\caption{
{\bf (a):} H30$\alpha$ emission map observed in ALMA C3+C6+C9
configuration, integrated in the velocity range $-10<\vlsr<+130~\kms$
(color scale and black contours at $5\sigma\times2^n$, $n=1,2,3,...$,
$1\sigma=500~\K~\kms$ or $0.036~\Jybeam~\kms$), overlaid with the 1.3
mm continuum in white contours.  
The dashed lines in panels indicate the
outflow axis ($\pa=3^\circ$) and projected opening angle ($40^\circ$).
{\bf (b):} Map of the integrated
H30$\alpha$ line-to-continuum ratio ($\int_v I_{\mathrm{H30}\alpha}
dv/I_\mathrm{1.3~mm}$, color scale), overlaid with the continuum
(black contours).
{\bf (c):} H30$\alpha$ and He30$\alpha$ spectra (solid black line) at the continuum
peak position.  The $3\sigma$ noise is marked by the error bar in each
velocity channel.  The red solid curve is the fitted profile combining
multiple Gaussian components, each of which is shown by the dashed
curves with their central velocities indicated by the dotted lines.
The inset panel shows a zoom-in view of the He30$\alpha$ spectrum.
{\bf (d):} The residual between the observed spectrum and the fitted
profile.  The shaded region indicates the 3$\sigma$ noise level.
{\bf (e):} Same as panel a, but showing the He30$\alpha$ emission
integrated in the velocity range $-150<V_{\mathrm{lsr,H30}\alpha}<-20~\kms$
($-28<V_{\mathrm{lsr,He30}\alpha}<+102~\kms$).
The black contours are at levels of $5\sigma\times2^n$, $n=1,2,3,...$,
$1\sigma=460~\K~\kms$ or $0.033~\Jybeam~\kms$). 
{\bf (f):} Same as panel d, but showing the integrated
He30$\alpha$ line-to-continuum ratio ($\int_v I_{\mathrm{He30}\alpha}
dv/I_\mathrm{1.3~mm}$).}
\label{fig:H30alpha}
\end{center}
\end{figure*}

The position and morphology of the H30$\alpha$ emission are found to
coincide very well with the 1.3 and 7~mm continuum emissions (Figures
\ref{fig:H30alpha}a and \ref{fig:H30alpha}b).  Figure \ref{fig:H30alpha}c shows the
H30$\alpha$ spectrum at the 1.3~mm continuum peak.  The H30$\alpha$
line has a complicated spectral profile, suggesting strong non-LTE
effects and multiple dynamical components.
In addition to the
strong narrow feature at $\vlsr=+33~\kms$ ($\FWHM=7~\kms$), the rest of the line profile
can be relatively well fit with three Gaussian components at
$\vlsr=+25$, $+44$ and $+80~\kms$ with $\FWHM=36$, 42, and $30~\kms$
(the source systemic velocity is $\vsys\sim +60~\kms$; \citealt[]{Ortega12}).  

To quantify the level of non-LTE emission, in Figure \ref{fig:H30alpha}b
we show the map of the integrated line-to-continuum (ILTC) intensity
ratio, $\int_v I_{\mathrm{H30}\alpha} dv/I_\mathrm{1.3~mm}$, where
$I_{\mathrm{H30}\alpha}$ is the H30$\alpha$ intensity and
$I_\mathrm{1.3~mm}$ is the 1.3~mm continuum intensity.  Under LTE
conditions and assuming both optically thin free-free
and H30$\alpha$ emissions, the ILTC intensity ratio is (see, e.g.,
eqs. 10.35, 14.27 and 14.29 of \citealt{Wilson13})
\begin{eqnarray}
&& \frac{\int_v I_{\mathrm{H30}\alpha} dv}{I_\mathrm{1.3~mm}} 
= 4.678\times 10^6~\kms \left(\frac{T_\mathrm{e}}{\K}\right)^{-1} \nonumber\\
&&\times\left[1.5\ln\left(\frac{T_\mathrm{e}}{\K}\right)-8.443\right]^{-1} 
\left[1+\frac{N\left(\mathrm{He}^+\right)}{N\left(\mathrm{H}^+\right)}\right]^{-1}.\label{eq:ILTC}
\end{eqnarray}
The last term results from electrons from He$^+$ contributing to
free-free emission but not to the HRL, and typically
$N\left(\mathrm{He}^+\right)/N\left(\mathrm{H}^+\right)=0.08$.
Assuming a characteristic ionized gas temperature of
$T_\mathrm{e}=10^4\:\K$, $\int_v I_{\mathrm{H30}\alpha}
dv/I_\mathrm{1.3~mm}=81\:\kms$, which is a reference value for the
optically-thin LTE conditions. As Figure \ref{fig:H30alpha}b shows,
most of the H30$\alpha$ emission is stronger than expected under
LTE conditions.  In some parts, the measured ILTC ratio is
$>200~\kms$ and reaching $\sim300~\kms$, significantly higher than the
LTE value of $81~\kms$, indicating strong maser amplification.
Previously, millimeter HRL masers were detected in
MWC349A, with an H30$\alpha$ ILTC ratio of $298\:\kms$
(\citealt[]{Martin89,Jimenez13}), and also in H26$\alpha$ (weakly in
H30$\alpha$) in MonR2-IRS2 (\citealt[]{Jimenez13}).  To our knowledge,
G45 is only the third massive young stellar object to show strong millimeter HRL masers,
with these reaching a similar level as in MWC349A.  Figure
\ref{fig:H30alpha}b also shows that the strong maser effect is
concentrated toward the east and west of the central source along a
direction nearly perpendicular to the outflow axis, i.e., along the
disk plane.  This behavior is similar to that seen in MWC349A, in
which HRL masers are found to be along certain annuli of the disk
(e.g., \citealt[]{Baez13,Zhang17}).
In addition, there are two small regions along the outflow cavity walls
to the north-east and north-west which also have strong maser levels.

Note that here we have assumed all the compact 1.3~mm continuum is free-free emission,
which is reasonably valid, as discussed above.  However, if there is a
significant dust contribution in the 1.3~mm continuum, the observed
ILTC ratio is then underestimated.  Furthermore, if the H30$\alpha$
emission is not completely optically thin, the expected ILTC ratio
under LTE conditions should be lower than the reference value of
$81~\kms$.  Both of these effects would imply even higher levels of
maser amplification.  In addition, the ionized gas temperature can
also affect the LTE value of ILTC ratio, which ranges from $34~\kms$ with
$T_\mathrm{e}=2\times10^4~\K$ to $156~\kms$ with
$T_\mathrm{e}=6\times10^3~\K$.  Even at the low temperature end of
this range, maser amplification is needed to explain the observed ILTC
ratios.

In addition to the H30$\alpha$ line, the He30$\alpha$ Helium
recombination line (231.99543~GHz) is also detected (Figure
\ref{fig:H30alpha}c).  Its spectrum can be fit with a single Gaussian
profile at $V_{\mathrm{lsr,H30}\alpha}=-65~\kms$
($V_{\mathrm{lsr,He30}\alpha}=+57~\kms$) and $\FWHM=83~\kms$.  This
line width is consistent with the total H30$\alpha$ line width.
He30$\alpha$ integrated emission and ILTC ratio maps are shown in
Figures \ref{fig:H30alpha}e and \ref{fig:H30alpha}f.  The He30$\alpha$
ILTC ratio increases from $\sim 5~\kms$ at the center to $\sim
20~\kms$ in the outer region.  The increasing ILTC ratio towards
outside is unexpected, as the outer region should have lower ratio
between Helium and Hydrogen ionizing photon rates.  One possible
explanation is that, the He30$\alpha$ emission also has maser effects
with a similar level to H30$\alpha$ (the expected He30$\alpha$ ILTC
ratio in optically thin LTE conditions is $6~\kms$ using similar
formula as Eq. \ref{eq:ILTC} and $T_e=10^4~\K$;
note that He$^{++}$ is not considered), but at the center,
the emission becomes partially optically thick, and the observed ILTC
ratio decreases.
Note that at $V_{\mathrm{lsr,H30}\alpha}\approx-50~\kms$, 
the dips are due to the absorption of CH$_3$OCHO $20_{9,12}-20_{8,13}$ line (231.9854~GHz)
and CH$_3$OCH$_3$ $13_{0,13}-12_{1,12}$ line (231.9879~GHz) from
the cooler surrounding material in the foreground.

\section{Discussions}
\label{sec:discussions}

\begin{figure*}
\begin{center}
\includegraphics[width=0.9\textwidth]{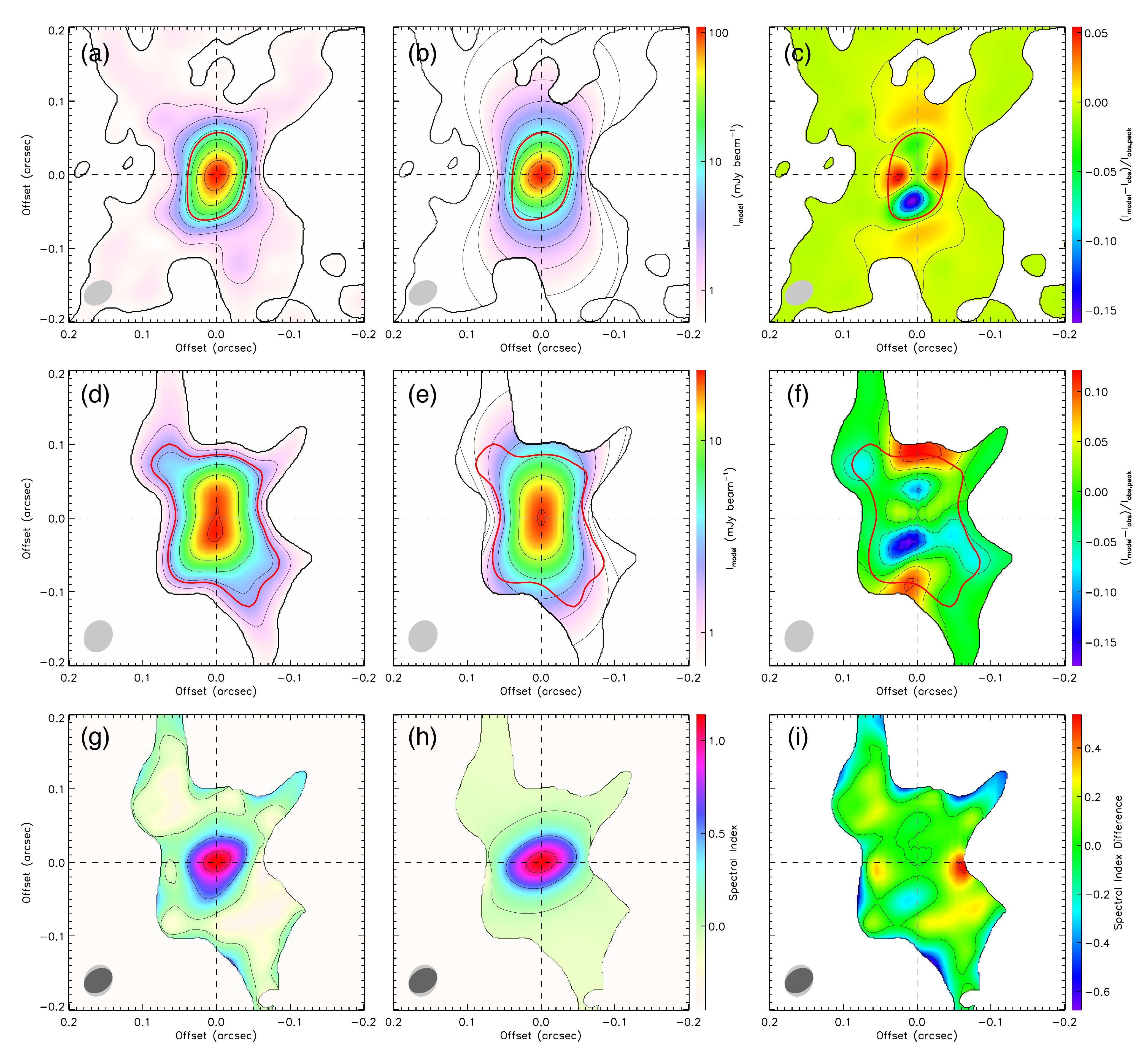}\\
\caption{
Comparisons between the model and observations in 1.3~mm continuum,
7~mm continuum and spectral index.  {\bf (a):} Observed 1.3~mm
continuum map.  {\bf (b):} 1.3~mm continuum map from the best-fit
model.  {\bf (c):} Differences between the observation and model
$(I_\mathrm{obs}-I_\mathrm{mod})/I_\mathrm{obs,peak}$.
{\bf (d)$-$(f):} Same as panels a$-$c, but for the 7~mm continuum.
{\bf (g)$-$(i):} Same as panels a$-$c, but for the spectral index map.
Only the regions with observed emissions $>10\sigma$ are shown.  The
red contours in panels a$-$f show the region with observed intensities
$>0.1I_\mathrm{obs,center}$, which is used in the 2D model fitting.  The maps
are rotated by $3^\circ$ so that the outflow axis is along the
$y$-axis.  The 7~mm observational image is shifted by offsets of 7 mas
in R.A. and 13 mas in Dec. to maximize the symmetry of the intensity
profiles with respect to the center.}
\label{fig:ffmodel}
\end{center}
\end{figure*}

\begin{figure*}
\begin{center}
\includegraphics[width=0.7\textwidth]{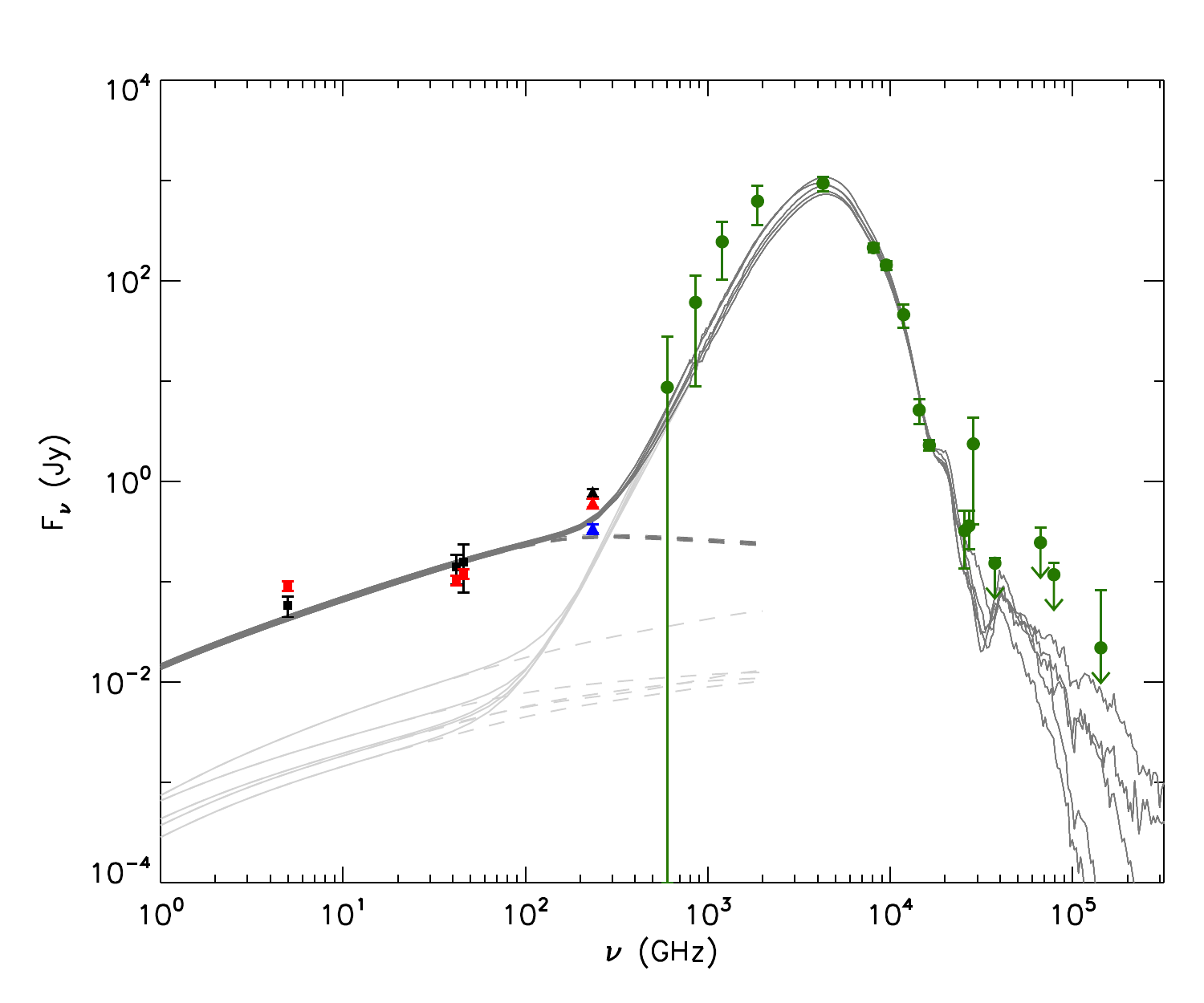}\\
\caption{
Observed SED of G45 compared to models.  Green symbols and error bars
show the fluxes from near-infrared to sub-mm wavelengths
(\citealt[]{Debuizer17}). Squares and error bars show the fluxes at 5,
42 and 46 GHz observed by VLA (\citealt[]{Rosero19}). The red and
black squares show the VLA fluxes integrated within $0.3\arcsec$ and
$14\arcsec$ from the center, respectively.  Triangles and error bars
show the fluxes at ALMA Band 6 with assumed 10\% error.
The black triangle shows the ALMA flux observed in C3+C6 configuration
and integrated within $5\arcsec$ from the center.  The red and blue
triangles show the ALMA fluxes within $0.9\arcsec$ from the center,
observed in C3+C6+C9 and C9 configurations, respectively.  The solid
lines show the combined SEDs from free-free models and dust continuum
models, and the dashed lines show the free-free model SEDs.  The light
dashed lines show the free-free SED from the model presented in
\citet[]{Rosero19}, based on the photoionized MHD disk wind model by
\citet[]{Tanaka16}.  The dark dashed lines shows the free-free SEDs
from the model presented in this work integrated over a region 
within $20000~\au$ (2.4$\arcsec$) from the center.  Best-fit models for different
opening angles ($\theta_w$) are shown but the differences are small.
The dust continuum models are from SED fitting by \citet[]{Debuizer17}
using the dust continuum radiative transfer model grid by \citet[]{ZT18}.}
\label{fig:sed}
\end{center}
\end{figure*}

\begin{figure*}
\begin{center}
\includegraphics[width=0.8\textwidth]{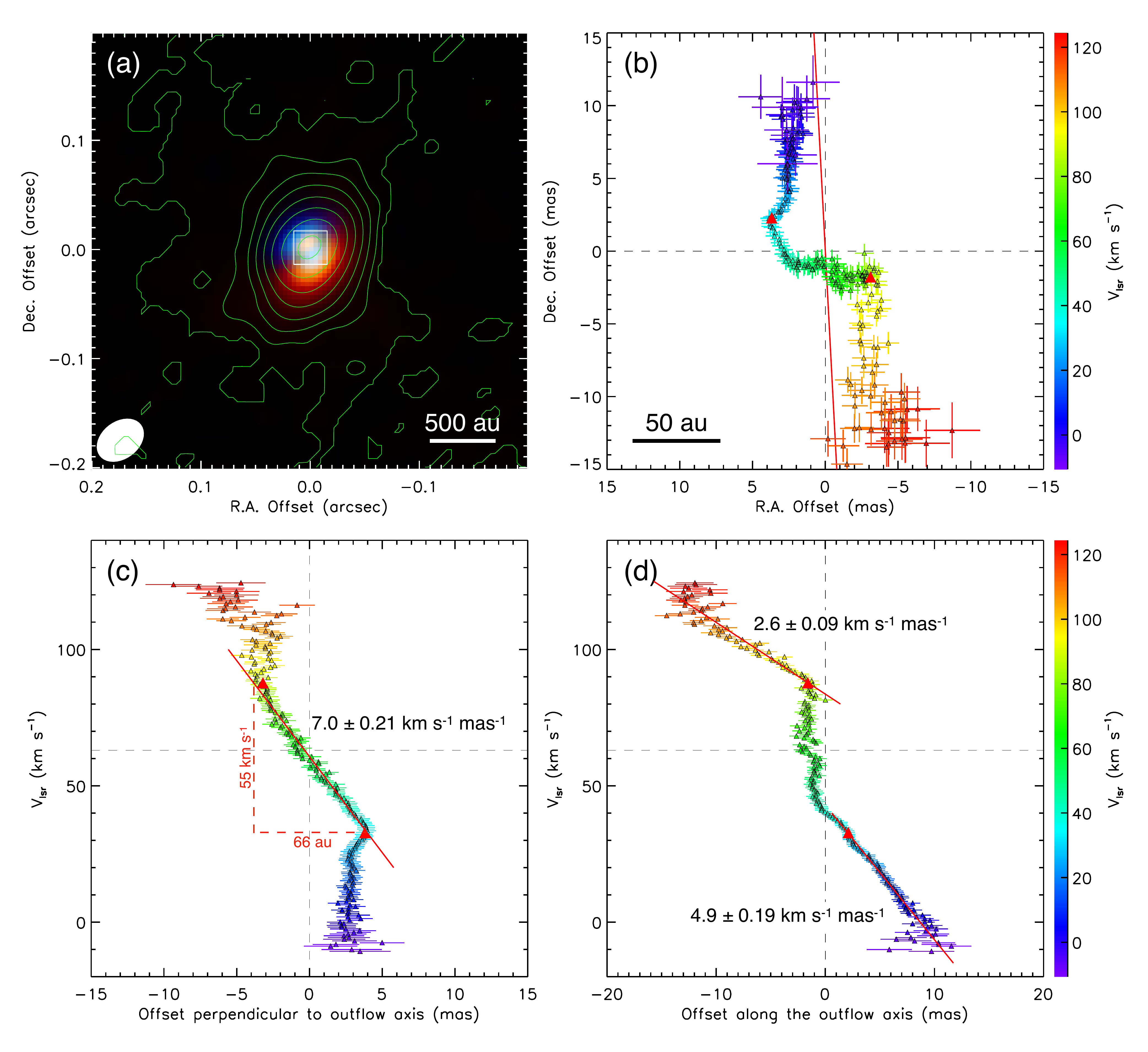}\\
\caption{
{\bf (a):} Integrated blueshifted ($-10<\vlsr<30~\kms$; blue color
scale) and redshifted ($90<\vlsr<130~\kms$; red color scale)
H30$\alpha$ emissions, overlaid with total integrated emission
($-10<\vlsr<130~\kms$; green contours; same as Figure
\ref{fig:H30alpha}a).  The white square marks the region shown in
panel b.  {\bf (b):} Distribution of the H30$\alpha$ emission
centroids (triangles with error bars).  Only channels with peak
intensities $>10\sigma$ ($1\sigma=1.8~\mJybeam$) are included. The
position offsets are relative to the continuum peak.  The
line-of-sight velocities are shown by the color scale.  The red line
indicates the outflow axis ($\pa=3^\circ$).  The large red triangles
mark the positions where the centroid distribution pattern changes.  
{\bf (c):} Distances of centroids from the continuum peak projected in
the direction perpendicular to the outflow axis with their
line-of-sight velocities. The red line indicates the fitted velocity
gradient perpendicular to the outflow axis.  {\bf (d):} Distances of
centroids from the continuum peak projected in the direction of the
outflow axis with their line-of-sight velocities. The red lines
indicate the fitted velocity gradients along the outflow axis
direction.}
\label{fig:centroid}
\end{center}
\end{figure*}

\subsection{Model for Free-free Emission}
\label{sec:model}

We construct a simple model to explain the 1.3 and 7~mm free-free
emissions, assuming isothermal ionized gas filling
bipolar outflow cavities. The details of the model fitting are
described in Appendix \ref{sec:appA}.  The best-fit models have
electron temperatures around $T_\mathrm{e}=1\times10^4~\K$, electron
number densities at the center around
$n_0=1.5\times10^7\:\mathrm{cm}^{-3}$ and outer radii of the outflow
launching region around $\varpi_0=110\:\au$.  The model constraints on
the true outflow opening angle (which depends on the inclination) are
weak.  Although the model with smallest $\chi^2$ has an outflow
half-opening angle of $\theta_w=22^\circ$, any $\theta_w$ value within
the range of $20^\circ$ to $40^\circ$ explored by the model grid can
generate good-fitting results with similar qualities (Appendix
\ref{sec:appA}).  Infrared SED fitting of this source provided a range
of inclinations of $i>55^\circ$ between the line of sight and outflow
axis, corresponding to $35^\circ<\theta_w<40^\circ$ (given projected
half-opening angle of $40^\circ$), which we consider to be a more
probable range for the opening angle.  The comparisons between the
observations and the best-fit model (minimum $\chi^2$) are shown in Figure
\ref{fig:ffmodel}.  The best-fit model successfully reproduces the
absolute values and relative distributions of both 1.3 and
7~mm intensities.  There are some emission excesses in the observation at
$\sim0.04\arcsec$ to the south of the center seen in both bands 
(stronger at 7~mm), which 
indicates some substructures in the ionized outflow and/or
contributions from nonthermal emissions, which are not taken into
account in the model. 
Although detailed modeling of the H30$\alpha$ emission
including the maser effects is out of the scope of this paper,
we note that the electron temperature and density distributions 
derived from the free-free model
are consistent with the observed strong H30$\alpha$ maser
(see Appendix \ref{sec:appA}).

The model requires an ionizing photon rate of $(1.4-1.8)\times
10^{49}~\mathrm{s}^{-1}$ (Appendix \ref{sec:appA}).  For zero-age
main sequence (ZAMS) stars, this corresponds to a stellar mass of 43
to $48\:M_\odot$ (\citealt[]{Davies11}), i.e., spectral type of
O4$-$O5 (\citealt[]{Mottram11}).  According to protostellar evolution
calculations with various accretion histories from different initial
and environmental conditions for massive star formation
(\citealt[]{Tanaka16,ZT18}), this ionizing photon rate
corresponds to a stellar mass of $28-48\:M_\odot$.  In all of these
protostellar evolution calculations, the central source has started
hydrogen burning and reached the main-sequence in the current stage.
These mass estimates are consistent with those estimated from infrared SED
fitting ($\sim20-50~M_\odot$, \citealt[]{Debuizer17}).  
%

Figure \ref{fig:sed} shows the observed SED from radio to
near-infrared. The infrared SED is well explained by dust continuum
emission (\citealt[]{Debuizer17}) using the continuum RT model
grid by \citet[]{ZT18}.  Based on the same physical model,
\citet[]{Rosero19} extended the model SED to radio wavelengths by
combining photoionization and free-free RT calculations by
\citet[]{Tanaka16} (thin lines in Figure \ref{fig:sed}).  However,
the observed radio fluxes are $\sim100\times$ higher than
these model predictions.  The new model presented here not only
reproduces the fluxes at 1.3 and 7~mm, but also simultaneously
reproduces the 1.3 cm fluxes (only low-resolution data is available at this wavelength).
Note that the new free-free model does not fit the SED directly, but rather, fit the
1.3 and 7~mm intensity maps.
At 1.3~mm, the free-free model can only reproduce the flux
of the compact emission (also see Appendix Figure \ref{fig:ffmodel_chisq}e), 
while the extended emission should be dominated
by the dust emission.

In the model presented in \citet[]{Rosero19}, 
only photoionization of the MHD disk wind was
considered, which has a typical density of $10^4\:\mathrm{cm}^{-3}$ in
the outflow cavity, while in the new model, a dense outflow with
densities of $\sim10^7~\mathrm{cm}^{-3}$ is included.  Such high
density suggests that this outflow is not an MHD disk wind, but
instead a photoevaporative outflow launched from the
disk.  The derived outflowing rate of this ionized outflow is 
$(2-3.5)\times10^{-5}(\vout/30\:\kms)\:M_\odot\:\yr^{-1}$ (Appendix
\ref{sec:appA}), consistent with theoretical expectations 
for photoevaporative outflow in later stage of massive star formation
(e.g.,\citealt[]{Tanaka17}).  Note that this outflowing rate is similar to or higher
than the typical outflow rates of MHD disk winds from massive protostars,
indicating that photoevaporation feedback is important at this stage.
However, MHD winds are still the dominant
feedback mechanism regulating core-to-star efficiency, as they
typically are much faster, and can
sweep up a much larger amount of ambient gas (\citealt[]{Tanaka17}).
Furthermore, in spite of the strong photoionization feedback, 
the hourglass morphology of the free-free emission suggests that
the ionized gas is still confined in the outflow cavity, while
the disk/envelope remains neutral along the disk mid-plane.

\subsection{Dynamics of the Ionized Gas}
\label{sec:dynamics}

Figure \ref{fig:centroid}a shows that the blue-shifted emission of
H30$\alpha$ is to the north of the continuum peak, and the red-shifted
emission is to the south, which indicates organized motion of ionized
gas.  To better demonstrate the kinematics of the ionized gas, we show
the H30$\alpha$ emission centroid distribution in Figure
\ref{fig:centroid}b (see Appendix \ref{sec:appB} for more details). 
The centroid distribution shows different patterns in different
velocity ranges.  At velocities $\vlsr\lesssim+33\:\kms$ and
$\vlsr\gtrsim+88\:\kms$, the centroids are distributed roughly along the
outflow axis direction, with more blue or redshifted centroids
further away from the center, consistent with outflowing motion with
acceleration.  At velocities $+33\:\kms\lesssim\vlsr\lesssim+88\:\kms$,
the centroids are distributed roughly perpendicular to the outflow
axis, with blueshifted centroids to the east and redshifted
centroids to the west, more consistent with rotation kinematics
(e.g., \citealt[]{Zhang19}).

These different patterns are better seen in panels c and d, in which
we plot the centroid offsets projected perpendicular to the outflow
direction and along the outflow direction separately.  In the velocity
range $+33\:\kms\lesssim\vlsr\lesssim+88\:\kms$, the centroids appear to
have a constant distance to the center in the direction of outflow
axis, but show a linear relation
between the velocity and offset in the direction perpendicular to the
outflow axis, which can be explained by rotation of a ring.  The
fitted velocity gradient within this velocity range is
$7.0\pm0.21~\kms~\mathrm{mas}^{-1}$ ($0.83\pm0.03~\kms~\au^{-1}$).  If
we consider the centroids of $\vlsr=+33$ and $+88~\kms$ are the two ends
of the projected ring (since they mark the changes in centroid
patterns; marked by red triangles), from their line-of-sight velocity
difference of $\Delta\vlsr=55~\kms$ and their position offsets
perpendicular to the outflow axis of $\Delta \varpi=66~\au$, we can
derive a dynamical mass of $m_\mathrm{dyn}=\Delta\vlsr^2\Delta
\varpi/(8G\sin^2 i)=28~M_\odot /\sin^2 i$.  For $55^\circ<i<90^\circ$
determined by the infrared SED fitting, we obtain
$m_\mathrm{dyn}=28-40~M_\odot$.  This mass estimate is consistent with
those derived from the free-free modeling ($28-48~M_\odot$) 
and infrared SED fitting  ($\sim20-50~M_\odot$). 

Along the outflow direction, the centroid velocity linearly increases
with distance at velocities $\vlsr\lesssim+33~\kms$ or
$\vlsr\gtrsim+88~\kms$ (marked by the red triangles).  The fitted
velocity gradients are $4.9\pm0.19~\kms~\mathrm{mas}^{-1}$
($0.58\pm0.02~\kms~\au^{-1}$) for the blue-shifted emissions and
$2.6\pm0.09~\kms~\mathrm{mas}^{-1}$ ($0.31\pm0.01~\kms~\au^{-1}$) for
the red-shifted emissions, which are among the highest seen in ionized
flows around forming massive stars (e.g., \citealt[]{Moscadelli18}).
Note that the two high-velocity Gaussian components in the H30$\alpha$
spectrum (Figure \ref{fig:H30alpha}c) have central velocities of $\vlsr=+25$ and
$+80~\kms$, which are close to the velocities where the centroid
pattern changes from rotation to outflow.  
FWHM of these two
components are 35 and 42~$\kms$, indicating motions with velocity
FWHM of 28 and 36~$\kms$ in addition to the thermal broadening with
a FWHM of $\sim20\:\kms$ for $\sim10^4\:\K$ ionized gas.
Therefore it is reasonable to consider that the ionized outflow has a
typical line-of-sight outflow velocity of $14-18~\kms$ in addition to
the rotation.  Considering projection effects, we adopt a fiducial
value of $\vout=30\:\kms$ for the velocity of this outflow
(\S\ref{sec:model}). This outflowing velocity is of order of the sound
speed ($10~\kms$) and thus consistent with the scenario of a photoevaporative
outflow (\citealt[]{hol94}).
A model including non-LTE effects is
needed to fully understand the H30$\alpha$ kinematics including its
transition from rotation to outflowing motion, which we defer to a
future paper.

\subsection{A Possible Embedded Jet}
\label{sec:jet}

As discussed in \S\ref{sec:continuum}, there is a jet-like structure
inside the wide-angle ionized outflow seen in the 1.3 mm continuum
morphology, along which several regions with negative spectral indices
are seen, indicating possible contributions from non-thermal
synchrotron emission.  Synchrotron emission is sometimes found
associated with high-velocity jets from massive protostars
(e.g., \citealt[]{Garay96,Carrasco10,Moscadelli13,Rosero16,Beltran16,Sanna19}).
\citet[]{Padovani15,Padovani16} performed model calculations
of particle acceleration in
protostellar jets under various conditions.  According to their model,
strong particle acceleration can happen under the appropriate
conditions for the G45 photoevaporative outflow
($\ga10^6~\mathrm{cm}^{-3}$ with high ionization fraction), if the
magnetic field strength is $\gtrsim1\:\mathrm{mG}$ and shock
velocities are $>100\:\kms$.
This high velocity needed for the particle acceleration suggests that the shock is
caused by a fast jet rather than the photoevaporation outflow itself ($\vout\sim30\:\kms$).
Fast jets are commonly considered as a strong indicator of on-going
accretion. Therefore this massive young star may be still
accreting, in spite of strong feedback from its own photoionizing
radiation.  This provides a direct observational confirmation that
photoionization feedback is not stopping accretion and limiting
the final mass even for protostars with masses $\sim50\:M_\odot$.

For the regions with negative spectral indices $<-0.3$ at
$\sim0.1\arcsec$ to the north and south of the central source, we
further estimate minimum-energy magnetic field strengths of
$B_\mathrm{min}=15$ and 17 mG for the synchrotron sources (see
Appendix \ref{sec:appC}).  Such values are much larger than those
estimated previously in synchrotron emitting regions around massive
protostars (e.g., $\sim1\:\mathrm{mG}$ in G035.02+0.35
(\citealt[]{Sanna19}) and $0.2\:\mathrm{mG}$ in HH80-81
(\citealt[]{Carrasco10})).  However, previously the synchrotron
emission was detected further away from the protostar (i.e.,
$10^4\:$au in G035.02+0.35 and $10^5\:$au in HH 80-81), while the
synchrotron emission is detected $<1000\:$au from the central source
in G45.  This magnetic field strength is consistent with some
observations.  For example, $\sim20\:\mathrm{mG}$ was measured in Cep
A HW2 within 1000 au (\citealt[]{Vlemmings10}).  It is also consistent
with the magnetic field predicted for the base of the outflow by
simulations of collapse of magnetically supported massive cores
(e.g., \citealt[]{Matsushita17,Staff19}).

\section{Summary}
\label{sec:summary}

With high-resolution ALMA and VLA observations at 1.3 and 7~mm, we
have discovered a bipolar wide-angle ionized outflow from the massive
protostar G45.47+0.05. The H30$\alpha$ recombination line shows strong
maser amplification, with this source being one of the very few massive protostars
so far known to show such characteristics. By modeling the 1.3 and 7~mm
free-free continuum, the ionized outflow is found to be a
photoevaporation flow launched from a disk of radius of $110~\au$ with
an electron temperature of $\sim10^4\:\K$ and an electron number density of
$\sim1.5\times10^7\:\mathrm{cm}^{-1}$ at the center.  The mass of the protostar is
estimated to be $\sim30-50\:M_\odot$ based on the required ionizing photon rate, 
or $\sim30-40\:M_\odot$ based on the H30$\alpha$ kinematics.  The
kinematics of the photoevaporated gas is dominated by rotation close
to the disk plane and outflow motion away from this plane. The derived
photoevaporative outflow rate is
$\sim(2-3.5)\times10^{-5}\:M_\odot\:\yr^{-1}$.  
With robust detection of a resolved bipolar photoevaporative outflow,
G45 provides a prototype of photoevaporation outflow in the later
stage of massive star formation.  Previously, clearly resolved bipolar
ionized structure was detected in MWC 349A, however, its evolutionary
stage or mass are still uncertain
(e.g., \citealt[]{Hofmann02,Baez13,Zhang17}).
G45 may also contain an inner, accretion-driven jet.  The possible
co-existence of a jet and a massive photoevaporative outflow suggests
that, in spite of strong photoionization feedback, accretion is still
proceeding to this massive young star. This confirms theoretical
expectation that radiation feedback plays a relatively minor role in
terminating accretion and determining the final mass of a forming
massive star.  The observed highly symmetric and ordered outflow and
rotational motions also suggest that this massive star is forming via
Core Accretion, i.e., in a similar way as low-mass stars.

\acknowledgements We thank Boy Lankhaar for valuable discussions.
This paper makes use of the following ALMA data:
ADS/JAO.ALMA\#2015.1.01454.S, and ADS/JAO. ALMA\#2017.1.00181.S.  ALMA
is a partnership of ESO (representing its member states), NSF (USA)
and NINS (Japan), together with NRC (Canada), MOST and ASIAA (Taiwan),
and KASI (Republic of Korea), in cooperation with the Republic of
Chile.  The Joint ALMA Observatory is operated by ESO, AUI/NRAO and
NAOJ.  Y.Z. acknowledges supports from RIKEN Special Postdoctoral
Researcher Program and JSPS KAKENHI grant JP19K14774.
K.E.I.T. acknowledges support from NAOJ ALMA Scientific Research Grant
Number 2017-05A, JSPS KAKENHI Grant Numbers JP19H05080, JP19K14760.
J.C.T. acknowledges support from ERC project MSTAR.
G.G. acknowledges support from CONICYT  project AFB-170002.

\software{CASA (http://casa.nrao.edu; \citealt[]{McMullin07}), 
The IDL Astronomy User's Library (https:// idlastro.gsfc.nasa.gov)}

\appendix
\section{Model Fitting of 1.3 and 7 mm Free-free Emissions}
\label{sec:appA}

We construct a simple model to explain the 1.3 and 7~mm free-free
emissions.  We assume isothermal ionized gas filling axisymmetric
bipolar outflow cavities that consist of two connected cones with
shapes described by
\begin{equation}
\varpi_{\rm out}(z)=\varpi_0+|z|\tan\theta_w,
\end{equation}
where $\varpi_{\rm out}$ is the outer radius of the outflow cavity
from the axis, $z$ is the height above the disk plane on both sides,
$\theta_w$ is the half-opening angle of the outflow cavity, and
$\varpi_0$ is the outer radius of the outflow launching region.  The
relation between the true outflow opening angle $\theta_w$ and the
projected opening angle $\theta_{w,\mathrm{sky}}$ is
\begin{equation}
\frac{\tan\theta_w}{\tan\theta_{w,\mathrm{sky}}}=\sin i, \label{eq:inc}
\end{equation}
where $i$ is the inclination angle between the outflow axis and the
line of sight.  For simplicity, we fix
$\theta_{w,\mathrm{sky}}=40^\circ$ based on the observed 1.3~mm
continuum morphology (Figure \ref{fig:contmap}).  The electron density
distribution of the ionized gas in the outflow is described by a power
law
\begin{equation}
n=n_0\left(\frac{r'}{z_0}\right)^{-2},
\end{equation}
where $r' = \sqrt{(|z|+z_0)^2+\varpi^2}$, $z_0=\varpi_0/\tan\theta_w$,
and $\varpi$ is the radius from the axis.  Here $r'$ is used to avoid a
singularity, rather than $r=\sqrt{z^2+\varpi^2}$.  In such a model, we
have four free parameters: electron temperature $T_\mathrm{e}$,
electron density at the center $n_0$, outer radius of the outflow
launching region $\varpi_0$, and outflow half-opening angle
$\theta_w$.  Free-free RT calculations (\citealt[]{Tanaka16}) are
performed to produce model images at 1.3 and 7 mm.
Here we only consider a situation that each ion has only one charge (1$e$),
such as H$^+$, He$^+$, i.e. not considering ions such as He$^{++}$.

We compare the model with observation by calculating
\begin{equation}
\chi^2=\int\left(\frac{\Imod-\Iobs}{\Iobs}\right)^2 d\Omega \Bigg/ \int d\Omega
\end{equation}
for 1.3 and 7~mm images respectively.  Here the integration is over a
region with $\Iobs>0.1 I_\mathrm{obs,center}$ at each wavelength.  To
achieve the best-fit model, we minimize $\chi^2_\mathrm{tot}\equiv
(\chi^2_\mathrm{1.3mm}+\chi^2_\mathrm{7mm})/2$ by varying the free
parameters within reasonable ranges of $5\times10^3~\K\leq
T_\mathrm{e}\leq 2\times10^4~\K$, $3\times 10^6~\mathrm{cm}^{-3}\leq
n_0 \leq 3\times 10^8~\mathrm{cm}^{-3}$,
$30~\au\leq\varpi_0\leq300~\au$, and $20^\circ\leq\theta_w\leq
40^\circ$.  Note that, in the observed data, the 1.3~mm emission peak
is slightly offset from either the emission peak or the center of
symmetry of the 7~mm image.  Therefore, we manually shift the 7~mm
image by 7 mas in R.A. and 13 mas in Decl.  so that the center of
symmetry of the 7~mm image coincides with the 1.3~mm emission peak (see Figure \ref{fig:ffmodel}d).
We also adopt a position angle of $3^\circ$ for the outflow axis.

\begin{figure*}
\begin{center}
\includegraphics[width=\textwidth]{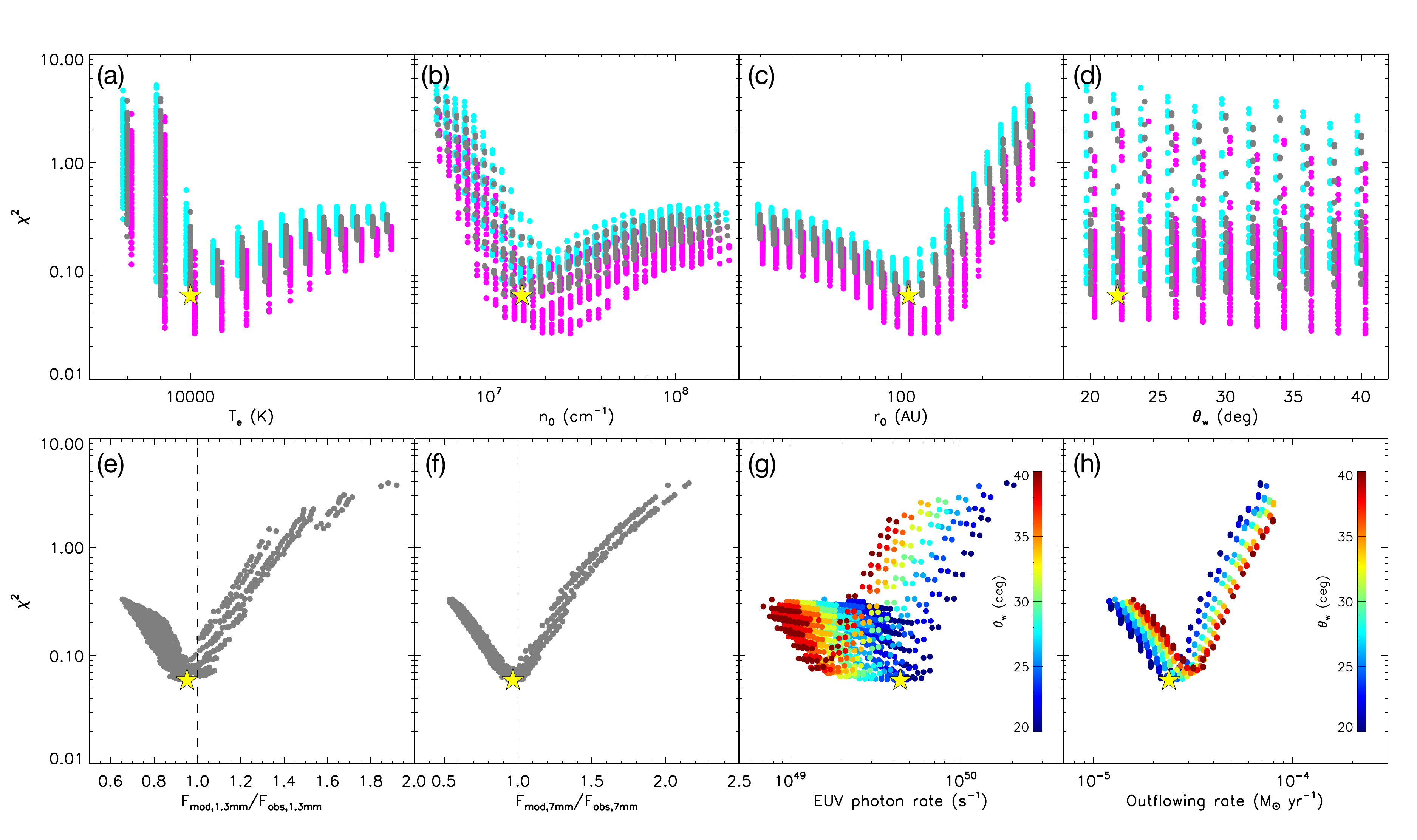}\\
\caption{
Distributions of $\chi^2$ values of free-free model fitting with model
parameters: electron temperature $T_\mathrm{e}$ (panel a), central
electron number density $n_0$ (panel b), radius of outflow launching region
$\varpi_0$ (panel c), outflow half-opening angle $\theta_w$ (panel d),
ratios of model and observed fluxes in the two bands (panels e and f),
ionizing EUV photon rate (panel g), and outflowing rate assuming a
velocity of $30~\kms$  (panel h).
In panels a$-$f, cyan, magenta and grey symbols show the
distributions of $\chi^2_\mathrm{7mm}$, $\chi^2_\mathrm{1.3mm}$ and
$\chi^2_\mathrm{tot}=(\chi^2_\mathrm{7mm}+\chi^2_\mathrm{1.3mm})/2$ of
the models, respectively.
In panels a$-$d, the symbols are slightly offset in x-axis for better presentation.
In panels e and f, the fluxes are integrated over the region with intensities 
$>0.1I_\mathrm{obs,center}$ (same as the region used in the 2D model fitting;
see red contours in Figure \ref{fig:ffmodel}).
In panels g and h, only
$\chi^2_\mathrm{tot}$ values are shown with the colors showing the
half-opening angle parameter $\theta_w$. The yellow star marks the
position of the best-fit model (minimum $\chi^2_\mathrm{tot}$).}
\label{fig:ffmodel_chisq}
\end{center}
\end{figure*}

The Appendix Figure \ref{fig:ffmodel_chisq} shows the distributions of the
$\chi^2$ values with various parameters.  As shown in panels a$-$c,
the parameters $T_\mathrm{e}$, $n_0$, and $\varpi_0$ are well
constrained by the model fitting.  The best-fit model (minimum
$\chi^2_\mathrm{tot}$) has $T_\mathrm{e}\approx10,000\:\K$,
$n_0\approx1.5\times 10^7~\mathrm{cm}^{-3}$ and
$\varpi_0\approx110~\au$.  Around these values both
$\chi^2_\mathrm{1.3mm}$ and $\chi^2_\mathrm{7mm}$ are close to their
minimum, showing that the model has simultaneous good fits to the
images of both bands.  However, the outflow half-opening angle
$\theta_w$ is not well constrained, as shown in panel d.  Although
$\theta_w\approx 22^\circ$ gives the minimum $\chi^2_\mathrm{tot}$, it
corresponds to an inclination angle of $i=29^\circ$ between the line
of sight and outflow axis (near face-on), which does not agree well
with other observations.  For example, infrared SED fitting
(\citealt[]{Debuizer17}) estimated the source inclination to be
$i>55^\circ$, corresponding to $35^\circ<\theta_w<40^\circ$, which we
consider the more probable range for the opening angle.  As the
constraint on opening angle (and inclination) is weak, the best-fit
model (minimum $\chi^2_\mathrm{tot}$, marked by a star in Figure
\ref{fig:ffmodel_chisq}) should be considered only as an example model
among a group of models with good fits to the observations.

\renewcommand{\figurename}{Appendix Figure}

\begin{figure*}
\begin{center}
\includegraphics[width=0.6\textwidth]{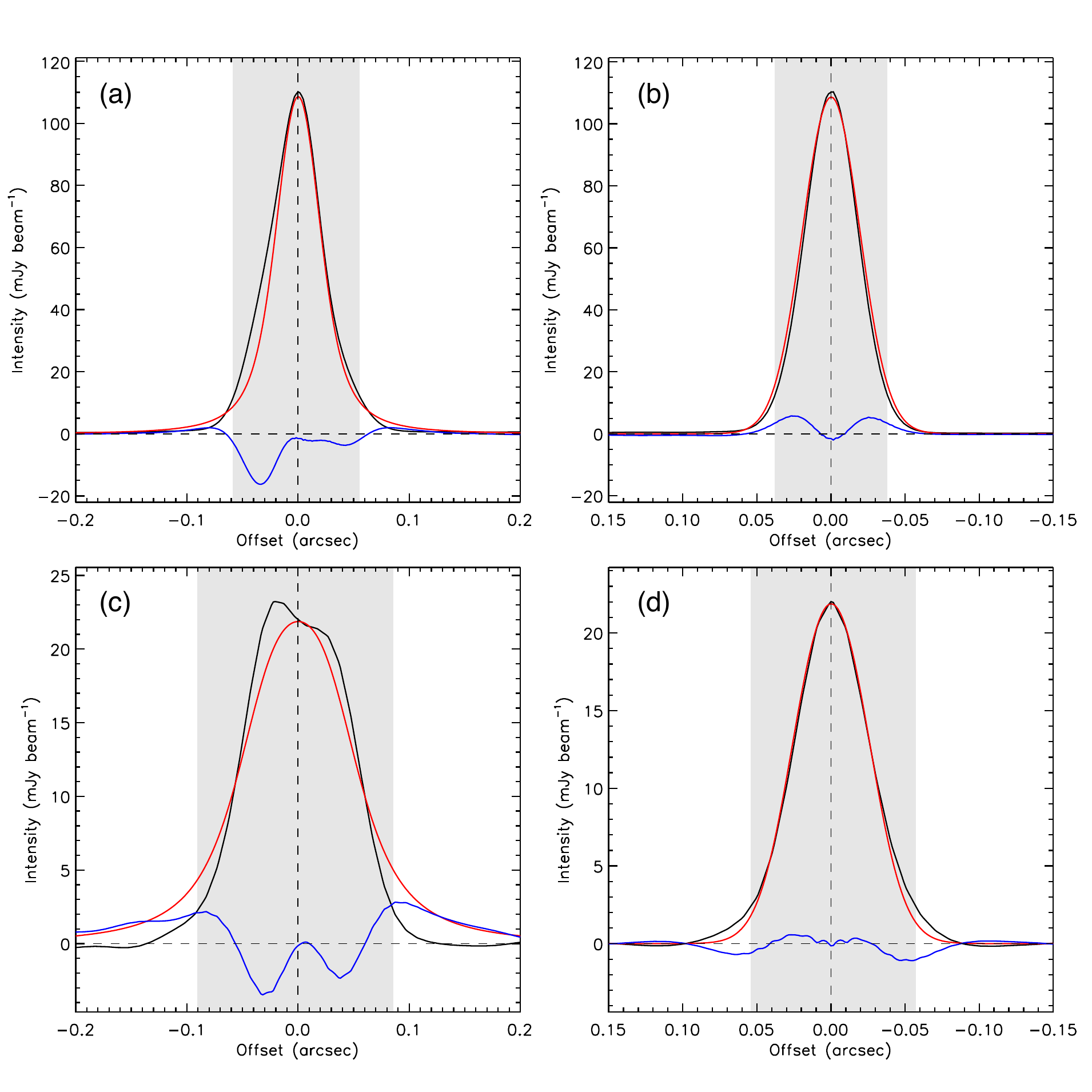}\\
\caption{Comparison of the observed continuum intensity profiles with the
model.  {\bf (a):} The observed 1.3~mm intensity profile along the
outflow axis (black) compared with the model (red), with the residual
shown in blue. The negative position offsets are to the south.  {\bf
  (b):} Same as panel a, but with the intensity profiles passing
through the center and perpendicular to the outflow axis. The negative
position offsets are to the west.  {\bf (c)$-$(d):} Same as panels
a$-$b, but for the 7~mm continuum.  The grey areas are with observed
intensities $>0.1I_{\rm obs,center}$, which is used in the 2D fitting.}
\label{fig:ffmodel_profile}
\end{center}
\end{figure*}

\begin{figure*}
\begin{center}
\includegraphics[width=0.6\textwidth]{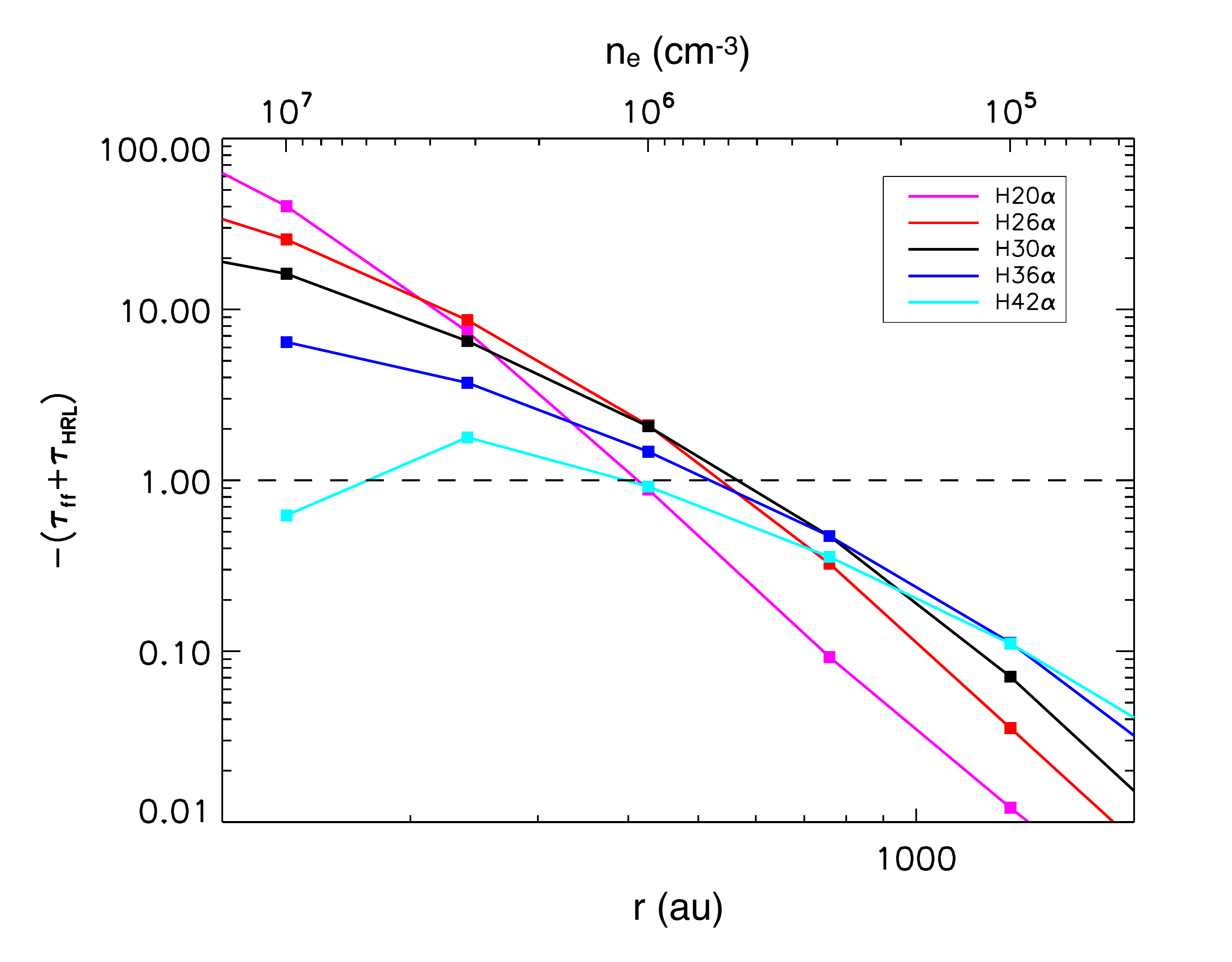}\\
\caption{Dependence of total continuum and hydrogen recombination line optical depths
$-(\tau_\mathrm{ff}+\tau_\mathrm{HRL})$ under non-LTE conditions
on the distance from the center or the electron density, 
estimated based on the free-free emission model.
Strong maser amplification requires $\tau_\mathrm{ff}+\tau_\mathrm{HRL} \ll -1$.
Curves with different colors show different hydrogen recombination lines.}
\label{fig:maser}
\end{center}
\end{figure*}

In addition to constraining the four free parameters, we also
calculate the total fluxes, ionizing photon rates and mass outflow rates from the
models, which are shown in panels e$-$h. 
Panels e and f show the comparison between the model and observation fluxes
at 1.3 and 7~mm integrated in regions with observed intensities 
$>0.1I_\mathrm{obs,center}$ (same as the region used in the 2D model fitting).
The model fluxes are slightly lower (within $5\%$) than the observed values.
Panel g shows the ionizing photon rate, which
is calculated following
\begin{equation}
S=4\pi\int_\mathrm{axis} \alpha_B n^2 z^2 dz,
\end{equation}
where $\alpha_B$ is the case B recombination rate for which we adopt
the fitting formulas from \citet[]{Draine11}, and the integration is
along the axis from 0 to $\infty$.  Note that this is only a lower
limit for the ionizing photon luminosity in the outflow,
which are just enough to ionize the outflow and ignore the possible effects of dust absorption.
The uncertainty of the EUV photon rate is relatively large. Models with
low $\chi^2$ values have EUV photon rates spanning from $\sim 1.4\times
10^{49}~\mathrm{s}^{-1}$ to $\sim 6\times 10^{49}~\mathrm{s}^{-1}$.
The uncertainty is mostly from the weak constraint on the opening
angle (and inclination), as shown in panel d.  The best-fit model
(minimum $\chi^2_\mathrm{tot}$) has an EUV photon rate of $4.5 \times
10^{49}~\mathrm{s}^{-1}$, while the most probable inclination range
($55^\circ<i<90^\circ$, i.e. $35^\circ<\theta_w<40^\circ$) gives a range $1.4 \times
10^{49}~\mathrm{s}^{-1}<S< 1.8 \times 10^{49}~\mathrm{s}^{-1}$.  Panel
h shows the mass outflow rates in the model (assuming a velocity of
$30~\kms$; see \S\ref{sec:dynamics}).  
The best-fit models have rates from $2\times 10^{-5}
(v/30~\kms)~M_\odot~\mathrm{yr}^{-1}$ to $3.5\times 10^{-5}
(v/30~\kms)~M_\odot~\mathrm{yr}^{-1}$.
These values are obtained assuming all the ions are H$^+$
and the total mass is 1.4 times of the hydrogen mass.
If we consider the He$^+$ contributions to the free-free emission,
the mass outflowing rates are $\sim7\%$ lower than the above values, 
assuming a constant ratio of 0.08 between He$^+$ and H$^+$
throughout the ionized region.

The comparisons between the observations and best-fit model are shown
in Figures \ref{fig:ffmodel} and \ref{fig:sed}, and the Appendix Figure
\ref{fig:ffmodel_profile}, including the images in both bands and
spectral index maps, the mm to radio SEDs, as well as intensity
profiles along and perpendicular to the outflow axis.  The best-fit
model successfully reproduces the absolute values and relative distributions of
emission intensities in both bands.  The profiles perpendicular to the
outflow axis are typically well fit, while some asymmetries or excesses
along the outflow axis still remain.  In particular, at $0.02\arcsec$
to the south, the observed 7~mm image shows an additional peak.  At
the same location, the observed 1.3~mm image also has a slight excess
(see the residual curves in panels a and c of Appendix Figure
\ref{fig:ffmodel_profile}).  This may be caused by some substructures
in the ionized outflow that are not taken into account in our model's
simple geometry and density distribution.  It is also possible that
this excess, stronger in the lower frequency, contains contributions
of non-thermal synchrotron emission.

Here we briefly discuss whether the observed H30$\alpha$ maser
can happen under the temperature and density conditions derived from
the free-free model.
Strong maser amplification requires the
total continuum and HRL optical depth
$\tau_\mathrm{ff}+\tau_\mathrm{HRL}\ll -1$ (\citealt[]{Baez13}).
The non-LTE HRL absorption coefficient
can be written as
$\kappa_\mathrm{HRL}=b_n\beta_{mn}\kappa_\mathrm{HRL,LTE}$,
with $\kappa_\mathrm{HRL,LTE}$ being the absorption coefficient
under LTE conditions, and $b_n$ and $\beta_{mn}$ being the
departure coefficients for the recombination line from electronic level $m$ to $n$
(e.g. $m=31,n=30$ for H30$\alpha$; \citealt[]{Dupree70}).
We adopt the values of $b_n$ and $\beta_{mn}$ derived by \citet[]{Walmsley90},
which are dependent on the electron temperature $T_e$ and density $n_e$. 
The non-LTE HRL optical depth is then
$\tau_\mathrm{HRL}=\kappa_\mathrm{HRL}l=b_n\beta_{mn}\kappa_\mathrm{HRL,LTE}l$,
with $l$ being the physical length scale.
Toward the center, with $n_e=n_0=1.5\times 10^7~\mathrm{cm}^{-3}$ and $T_e=10,000~\K$
from the free-free model, and assuming the length scale $l=\varpi_0=110~\au$ 
for a first-order estimation, we estimate $\tau_\mathrm{HRL}\approx -20$.
The total continuum and HRL optical depth is then $\tau_\mathrm{ff}+\tau_\mathrm{HRL}\approx -19$,
which is consistent with the observed strong maser amplification (Appendix Figure \ref{fig:maser}).
We further estimate the distribution of non-LTE HRL optical depth over the ionized region,
by adopting a simple relation $n_e\propto r^{-2}\sim l^{-2}$.
The dependence of  $\tau_\mathrm{ff}+\tau_\mathrm{HRL}$ on the distance
to the center $r$ or the density distribution $n_e$ is shown in the Appendix Figure \ref{fig:maser}.
Strong maser amplification appears constrained in the region $r\lesssim 400~\au$
for H30$\alpha$ line, which is consistent with the observation that the high
maser amplification is concentrated close to the disk mid-plane.
It also shows that similar maser effects may be seen in HRLs 
from about H20$\alpha$ to about H42$\alpha$,
which may be tested by future observations.
However, accurate model prediction for HRL maser intensities is
difficult as full consideration of the radiation field is need,
which is out of the scope of this paper.

\section{Determining the H30$\alpha$ emission positional centroids.}
\label{sec:appB}

The centroid positions are determined by fitting
Gaussian ellipses to the H30$\alpha$ emissions observed in the C9
configuration at channels with peak intensities $>10\sigma$.  The
accuracy of the centroid positions are affected by the signal-to-noise
ratio (S/N) of the data, following $\Delta
\theta_\mathrm{fit}=\theta_\mathrm{beam}/(2~\mathrm{S/N})$
(\citealt{Condon97}), where $\theta_\mathrm{beam}$ is the resolution
beam size, for which we adopt the major axis of the resolution beam
$\theta_\mathrm{beam}=38~\mathrm{mas}$ (320~au).  The phase noise in
the bandpass calibrator also introduces an additional error to the
centroid positions through passband calibrations (\citealt{Zhang17}).
The phase noise in the passband calibrator J2000-1748 is found to be
$\Delta \phi=5.4^\circ$ after smoothing of 4 channels. Such smoothing
is the same as that used in deriving the passband calibration
solutions.  The additional position error is $\Delta
\theta_\mathrm{bandpass}= \theta_\mathrm{beam}(\Delta\phi/360^\circ)$,
and the uncertainties in the centroid positions are $\Delta
\theta_\mathrm{centroid}=\sqrt{\Delta \theta_\mathrm{fit}^2+\Delta
  \theta_\mathrm{bandpass}^2}$.

\section{Estimating the Magnetic Field Strength from Synchrotron Emission}
\label{sec:appC}

In the region with negative spectral indices indicating synchrotron
emission, we estimate the minimum-energy magnetic filed strength,
$B_\mathrm{min}$, which minimizes the total energy of the synchrotron
source by assuming equipartition between the magnetic field energy and
the particle energy.  Following the classical formula for estimating
minimum-energy magnetic field strength
(e.g., \citealt[]{Carrasco10,Sanna19})
\begin{equation}
B_\mathrm{min} =[4.5 c_{12}(1+k)L_R]^{2/7}(R^3)^{-2/7}, \label{eq:sync}
\end{equation}
where $L_R$ is the synchrotron luminosity, $R$ is the source radius,
$k$ is the energy ratio between the ions (carrying most of the energy
but not radiating significantly) and electrons, and $c_{12}$ is a coefficient
dependent on the spectral index and the minimum and maximum
frequencies in the integration of the spectrum.  Here $B_\mathrm{min}$
is in Gauss, and $L_R$, $R$, $k$ and $c_{12}$ are in cgs units.  We
define the synchrotron emission region as the region with intensity
spectral indices of $\alpha<-0.3$ and both 1.3 and 7 mm emissions
$>5\sigma$.  Here we only consider the negative spectral index regions
within $\sim0.2\arcsec$ from the center, as the 7~mm image is affected
by strong sidelobe patterns further out from the central source.  The areas of
the regions to the north and south of the central source are
$0.00021\mathrm{arcsec}^2$ and $0.0013\mathrm{arcsec}^2$, which
correspond to effective radii of $R=0.008\arcsec$ and $0.020\arcsec$,
which we use as the radii of the synchrotron emitting regions.
Assuming all the 1.3 and 7 mm emissions from these regions are due to
synchrotron emission, we obtain synchrotron fluxes of 0.041 and 0.070
mJy at 1.3 and 7 mm in the northern region, and 1.0 and 1.7 mJy at 1.3
and 7 mm in the southern region.  Note that these are only upper
limits for the synchrotron fluxes, as these regions only have negative
spectral indices $\sim-0.3$ while typical synchrotron emission has a
spectral index of $\sim-0.75$, suggesting there are both free-free and
synchrotron contributions in these regions.  However, the magnetic
field strength is not very sensitive to the synchrotron flux (see
Eq. \ref{eq:sync}).  We then calculate synchrotron luminosities of
$L_R=8.1\times10^{29}~\mathrm{erg}~\mathrm{s}^{-1}$ and
$L_R=2.0\times10^{31}~\mathrm{erg}~\mathrm{s}^{-1}$ in the northern
and southern regions, by integrating the synchrotron fluxes from 1.3
to 7 mm assuming a power-law distribution.  For this frequency range,
we obtain $c_{12}=3.2\times 10^6$ following \citet[]{Govoni04}, We
also adopt $k=40$ following \citet[]{Carrasco10} and \citet[]{Sanna19}.  These
values give $B_\mathrm{min}=15$ and 17~mG in the northern and southern
regions, respectively.

\clearpage

\clearpage

\clearpage

\end{document}